\documentstyle[prd,aps,eqsecnum,tighten]{revtex}
\begin{document}
\draft
\title{Gauge Theory of Oriented Media. } 
\author{ Ivangoe.~B.~Pestov\thanks{Electronic address:
pestov@thsun1.jinr.ru}}
\address{Bogoliubov Laboratory of Theoretical Physics, Joint
Institute for
Nuclear Research \\ Dubna, 141980, Russia}

\date{\today}
\maketitle
\begin{abstract}
The concept of wave field is introduced  to represent oriented
media. The wave field is a tensor field of second rank, and
directors are its eigenvectors. This exhibition of
directors defines a natural gauge group inherit in continua and
allows one to derive from variational principle general relativistic
and gauge invariant equations for the wave field in question.  Thus,
the gauge-theoretical approach to continuum with internal degrees of
freedom gives unambiguous and minimally coupled theory.
\end{abstract}

\section{Introduction}
 In generalized continuum mechanics dealing with oriented media
 the fundamental structure is a set
 of vectors adjoined to earth point of  medium. These
 vectors which can deform independently of the displacements of
 points  of media are called directors. They define a microstructure
 of continuum. In the theory
 of spinning fluids, the name tetrads is also used. At present, this
 part of mechanical science is fairly well accepted and is widely
 applied for the description of media possessing not only energy -
 momentum but also internal degrees of freedom.

 There exists an extended literature on continua with
 internal degrees of freedom, see, for example, Eugene Cosserat and
 Francois Cosserat [1], Weyssenhoff and Raabe [2], Maugin and Eringen
 [3] and references herein, Halbwachs [4],
 Minkevich and Karakura [5], Berman [6].
 In the theory of elastic continua with defects the gauge approach
 was successfully used by Osipov [8].The theory of spinning fluid in
 generalized space-time manifolds was developed by Ray and Smally [7]
 as an extension of the theory of spinning fluids in special
 relativity.  The Cosserat brothers were first who introduced the
notion of a 3-tuple of unit rigid directors and laid down the
 mathematical foundations of the theory now a day known as the
 Cosserat continuum.

 The goal of present paper is to
 formulate the most general gauge-theoretical approach to the theory
 of oriented media.  To do this, the notion of a wave field is
 introduced to represent generalized  continua. The wave field is
 characterized so as to define directors and other properties of the
 system in question, in particular, the form of interaction of this
 matter with the gravitational field.  The wave field allows one to
 derive a minimally coupled and simple theory  from the first
 principles which have fundamental meaning in contemporary physics.

 \section{Formulation of the problem }
 The key idea of the present consideration is that directors are
 defined as  solutions of the eigenvalues problem of the form
    \begin{equation} \Psi^i_j h^j = \lambda h^i, \end{equation} where
 the matrix  $\Psi^i_j,$  is a tensor field of the type (1,1) and,
 respectively, the wave field of  oriented media.   The internal
 state of media is then characterized by the number of linear
 independent eigenvectors (directors) $h^i$ which define the type of
 microstructure of continua.  Now, the problem is to derive a wave
equation for the field $\Psi^i_j $ from the first principles.

We remark that equations (1) are invariant under the
 transformations \begin{equation} \bar \Psi^i_j = S^i_k \Psi^k_l
		T^l_j, \end{equation} \begin{equation} \bar h^i = S^i_j
	    h^j, \end{equation} where $S^i_j$ are components of the
 tensor field $S$ of type (1,1) that satisfy the condition  $ \det
 (S^i_j) \not= 0.$  In this case, there exists an inverse
 transformation  $S^{-1}$ with components $T^i_j$ such
 that $S^i_k \,T^k_j = \delta^i_j.$
 It is evident  that substitutions (2) form a local group
 of transformations, and equations for the field $\Psi$ should be
 invariant with respect to this gauge symmetry group. As it is
well-known, the matrix $\Psi$ by the gauge transformation can be
 reduced to the canonical or Jordan form that is defined by the
 characteristic equation
 $$|\Psi^i_j - \lambda \delta^i_j| = 0.$$

 The Einstein General Covariance is the other
 deep guiding principle that we have at our disposal.  In this
 framework, the wave equation for $\Psi$ is defined in fact uniquely.

The most direct and simplest way to derive gauge-invariant wave
 equation is  to give a correct definition of the gauge covariant
 derivative. Denote it by $D_i$ and let
 $$
	   D_i\Psi^j_k = \partial_i \Psi^j_k  + G^j_{il} \Psi^l_k  -
 \Psi^j_l G^l_{ik}. $$
 $$ \bar D_i \bar \Psi^j_k = \partial_i \bar \Psi^j_k  +\bar
G^j_{il} \bar \Psi^l_k - \bar \Psi^j_l \bar G^l_{ik}, $$  where
$G^j_{ik}$ and $\bar G^j_{ik}$ are  gauge potentials connected with
 wave fields $\Psi$ and  $\bar \Psi,$  respectively.

 For brevity, we will use the matrix notation
 $$\Psi = (\Psi^i_j), \quad G_i = (G^j_{ik}),\quad S=(S^i_j),
 \quad S^{-1} = (T^i_j), \quad  Tr(\Psi)=\Psi^i_i,$$
 in which \begin{equation}D_i\Psi = \partial_i\Psi + [G_i,
 \Psi],\quad \bar G_i \bar \Psi = \partial_i \bar \Psi + [\bar
 G_i,\bar \Psi ], \quad \bar \Psi = S  \Psi S^{-1}.    \end{equation}
 From the condition \begin{equation}\bar D_i \bar \Psi = S D_i \Psi
 S^{-1} \end{equation} and (4) one derives the law of transformation
 of the gauge potential \begin{equation} \bar G_i = S G_i S^{-1} + S
 \partial_i S^{-1}.  \end{equation} To derive general covariant
 Lagrangian of first order for $\Psi,$ it is natural to take $ D_i
 \Psi^j_k $ to be a tensor. Of course, the second gauge-covariant
 derivative should not be a tensor, but if we deal with the general
 covariant Lagrangian of first order, then by varying we will get a
 combination of gauge-covariant derivatives such that equation of
 second order will be tensor equation.

Let $$ \tilde
 x^i = \tilde x^i (x^0, x^1, x^2, x^3), \quad x^i = x^i (\tilde x^0,
 \tilde x^1, \tilde x^2, \tilde x^3)$$ coordinate transformation. Put
 $$X =(\frac{\partial \tilde x^i}{\partial x^j}), \quad X^{-1} =
 (\frac{\partial x^i }{\partial \tilde x^j}) ,$$ then
 the  gauge potential transforms as follows
 $$\tilde G_i = X G_k X^{-1} \frac{\partial x^k  }{\partial \tilde
x^i } + X \frac{\partial}{\partial \tilde x^i } X^{-1}.$$

Since  the tensor field $\Psi$  transforms under gauge and
coordinate transformations by the law $\bar \Psi = S \Psi S^{-1},
\quad \tilde \Psi = X \Psi X^{-1}, $ then the traces $\Psi^i_i ={\rm
Tr} \Psi $ and $\Psi^i_j \Psi^j_i = {\rm Tr}(\Psi\,\Psi)$  are
evidently invariants of the gauge group and scalar with respect to
the general coordinate transformations.  It is known from the theory
of linear operators that there also exist other invariants, but in
what follows we will only use these simplest ones.

   To get an expression for the tensor of strength  of the gauge
field, consider the commutator of covariant derivatives $[D_i,
\,D_j].$ From (4) it follows that
\begin{equation}
		[D_i,\,D_j]\Psi = [H_{ij},\, \Psi],
\end{equation}
where
\begin{equation}
  H_{ij}= \partial_i G_j -  \partial_j G_i  + [G_i,\,G_j]
\end{equation}
is the strength tensor of the gauge field with the following
properties
\begin{equation}
	 \bar H_{ij} = S H_{ij}S^{-1}, \quad  [D_i,\,D_j] H_{kl} =
[H_{ij},\,H_{kl}].
\end{equation}
 From (8) it follows that
$H_{ij}=(H_{ijl}^k) $  is a tensor of the type (1,3) with respect to
the general coordinate transformation.

\section{ Gauge-Invariant Equations}

Now we have all that is required to
derive the simplest general covariant and gauge-invariant equations
for fields $G_i$ and $ \Psi. $ In what follows $g_{ij}$ are the
Einstein gravitational potentials, $g^{ij}$ are components of tensor
inverse to $g_{ij}, \quad g_{il} g^{jl} = \delta^i_j.$ As it is
known, the determinant $|g_{ij}| \not= 0 ,$ which actually allows
one to obtain, for the tensor field $g_{ij},$ the equations invariant
under the general coordinate transformations.  By analogy, let us
consider the case when the determinant $|\Psi^i{_j} |\not= 0.$ Under
this condition the field $\Psi$ has the inverse one and the
nonlinear gauge-invariant equations can be suggested.  This means
that we consider the case of non-linear continuum mechanics with
internal degrees of freedom.  The linear case can be consider then as
an approximation.

Thus, the gauge-invariant and general covariant Lagrangian of first
order has the form \begin{equation} L= -\frac{a}{2} {\rm Tr}(D_i \Psi
D^i \Psi^{-1} ) -\frac{b}{4}{\rm Tr} (H_{ij} H^{ij} ), \end{equation}
where $a$ and $b$  are constants,  $$D^i = g^{ij} D_j \mbox{ and}
\quad H^{ij} =g^{ik} g^{jl} H_{kl}.$$ The gauge potential has
dimension $cm^{-1},$ $\Psi$ is dimensionless. Taking into account
that $$\delta \Psi= -\Psi(\delta \Psi^{-1} )\Psi $$ and varying (10)
with respect to $\Psi$ and $G_i, $ we obtain the following equations
of second order for basic fields  \begin{equation} D_ {i} (\sqrt{
|g|} \Psi^{-1} D^ {i} \Psi)\,=0, \end{equation}

\begin{equation} D_ {i} (\sqrt{ |g|} H^{ij})\,= \,\sqrt{
|g|} J^j , \end{equation} where $|g|$ is the absolute value of the
determinant of the matrix $(g_{ij})$ and $$ J^ {i}\, =\,[\Psi^{-1}
,D^ {i} \Psi] .$$ The tensor current $J$ has to satisfy the
equation $$ D_i (\sqrt{|g|}J^i )=0 $$ as in accordance with (9),
$$D_i D_j (\sqrt{ |g|}H^{ij})=0 .$$ Since this is really so, the
system of equations (11) and (12) is compatible.

Varying the
Lagrangian (10) with respect to $g_{ij},$   we obtain the
gauge-invariant metric tensor of energy-momentum of oriented media $$
T_ {ij} = a {\rm Tr}(D_ i \Psi D_ j \Psi^ {-1}
)+b {\rm Tr}(H_ {ik} H_ j^ k  )+g_ {ij}  L  $$ which satisfies
the equation \begin{equation} T^{ij}\,_{\!;i} = 0 .  \end{equation}
The semicolon denotes the covariant derivative with respect to the
Levi-Civita connexion belonging to the field $g_{ij}$ $$ \{^i_{jk}\}
= \frac{1}{2}g^{il} (\partial_j g_{kl} + \partial_k g_{jl} -
\partial_l g_{jk} ).$$ When deriving (13), besides equations of
fields, one should use the standard relations [9] for the
Christoffel symbols $\{^i_{jk}\} $ and the identity $$     D_i H_{jk}
+D_j H_{ki} +D_k H_{ij} =0   $$ which can easily be obtained with the
help of relation (7).  From (13) and the gauge invariance of the
 metric tensor of energy-momentum it follows that the Einstein
equations $$R_{ij} -\frac{1}{2} g_{ij} = \frac{G}{c^3} T_{ij},$$
derived from the Lagrangian $L_f \,= \,L_ g +L ,$ where $L_ g =
\frac{c^3}{G} R$ is the Einstein-Hilbert Lagrangian, will be
compatible. Thus, it is shown that continuum gravitates.

In the linear approximation $\Psi^i_j = \delta^i_j + M^i_j ,$  we
have from (11) the following equation for the matrix $M = (M^i_j)$
\begin{equation}D_{i} (\sqrt{ |g|}D^i M = 0.  \end{equation} We will
say that continuum admits elastic deformations if the vector field
$V^i$ exists such that \begin{equation} D_i V^j = 0.  \end{equation}
If equation (15) has a nontrivial solution, then from (14) and (15)
it follows that the vector field $U^i = M^i_k V^k$  obeys the equation
\begin{equation}
	   D_{i} (\sqrt{ |g|}D^iU^j) = 0
\end{equation}
that can be considered as a general covariant and gauge-covariant
analog of the known equation of elastic deformations of media.

Remark some more special cases of the general picture outlined.  The
consideration of rigid directors or generalized Cosserat
continua means, in the framework of the gauge approach, the reduction
of the gauge group defined by the condition
			$$S^i_k S^j_l g_{ij} = g_{kl}.$$

If we should like to set aside some director  then we have the
following constraint on the group of gauge transformations $$S^i_j
h^j = h^i.$$ At last, in absence of gravitational field we put
$g_{ij} ={\rm diag} (1,\,-1,\,-1,\,-1).$ We do not see other cases
to be mentioned.

Some very interesting gauge-invariant quantities can
be constructed from the fields $\Psi$ and $H_{ij} .$ We dwell here on
some of them.  The invariants, in particular, are $$ \varphi =
{\rm Tr}\Psi,\quad \triangle = |\Psi^i_j |, \quad F_{ij} = {\rm Tr}
H_{ij} .$$ If $\Psi$ obeys equation (11), then taking the trace of
both sides of this equation we obtain that the invariant  $\triangle
$ satisfies the equation $$ \partial_i (\sqrt{ |g|} g^{ij}\partial_j
ln|\triangle |)= 0 .$$
From (12) it follows that the bivector $F_{ij}$  satisfies the free
Maxwell equations
$$\partial_i (\sqrt{ |g|} F^{ij}) = 0.$$
Let $Q_i = trG_i = G^k_{ik}.$  According to (6) and the
 differentiation rule for determinants, the transformation law for
$Q_i$  under gauge transformations has the form
$$  \bar Q_i = Q_i - \partial_i ln|D|, $$
where $D={\rm det} (S^i_j).$  From (8) it follows that $F_{ij}
=\partial_i Q_j - \partial_j Q_i.$

A state of media
$(\Psi,H_{ij} ) $ is said to be singlet if it is invariant under all
the symmetry transformations.  In our case a singlet state is given
by the equations
$$ \Psi=S \Psi S^{-1} , \quad H_{ij} =S H_{ij} S^{-1} $$ to be satisfied at any $S.$ The first equation has the
solution $$\Psi=e^{\alpha}  E ,\quad \Psi^{-1} =e^{-\alpha} E ,$$
where $\alpha $ is a scalar field.  In this case all directions are
interchangeable. If the  gauge field obeys the equation
$H_{ij} =S H_{ij} S^{-1}, $ it also obeys the equation $H_{ij} = 1/4
F_{ij} \, E.$  Thus, a singlet state of media is represented by the
scalar field and 1-form $Q_i dx^i.$

\section{Conclusion} Outline the main principles of
gauge approaches to the continua with internal degrees od
freedom. In the proposed theory directors are not fundamental objects
and are treated as  solutions of the algebraic equation (1) at
given matrix $\Psi$  which is a wave field that corresponds to the
continuum in question and define all important quantities of oriented
media. The field $\Psi$ is deduced from equations (11) and (12). The
gravitational interactions of oriented media are described by the
Einstein equations with the energy-momentum tensor of continua given
by equation (13).  For equations presented, the gauge invariance
holds necessary valid in a sense that if $\Psi, \, G_i$  is a
solution then  $$\bar \Psi = S\Psi S^{-1},
\quad \bar G_i = SG_i S^{-1} + S \partial_i S^{-1}$$  is a
solution as well, where $S$ is any nondegenerate tensor field of the
type (1,1).  Such a general realization of gauge principle
presupposes not only direct application of the equations in question
but allows one to put forward the conjecture that oriented media
presented above are a new fundamental type of matter.  


\newpage

\begin{thebibliography}  {10}
\bibitem{1} Cosserat E. and  Cosserat F. 1909. {\it Th\'eorie des
corps d\'eformable} ( Hermann, Paris )
\bibitem{2} Weyssenhoff J. and Raabe A.  1947. {\it Acta Physica
Polonica},{\bf9} 7.
\bibitem{3}  Maugin G. A. and Eringen A. C. 1972 {\it
J.Math.Phys.}{\bf13} 1788.
\bibitem{4} Halbwache E.  1960.
{\it Th\'eorie relativiste des fluides \`a spin} ( Paris:
Gauthier-Villars)
\bibitem{5} Minkevich A. Y. and Karakura F.
1983 {\it J.Phys.} A {\bf 16} 1409.
\bibitem{6}  Berman M. S.  1990. {Nuov.Cim.} B {\bf 105}
235.
\bibitem{7} Osipov V. A. 1991. {\it
J.Phys.A:  Math.  Gen.} {\bf 24} 3237.
\bibitem{8}  Ray J. R. and
Smally L. L. 1982 {\it Phys.  Rev.} D {\bf 26} 2615.
\bibitem{9}  Schouten J. A. 1951. {\it Tensor Analysis for Physicists}
(Oxford: At the Clarendon Press )

\end{thebibliography}
\end{document}